\documentclass{elsart}
\usepackage{graphicx,amssymb}
\usepackage{subfigure}

\begin{document}
\begin{frontmatter}
\title{A new method of alpha ray measurement using a Quadrupole Mass Spectrometer} 

\author[Physics]{Y. Iwata\corauthref{cor1}},
\ead{yiwata@icepp.s.u-tokyo.ac.jp}
\author[ICEPP]{Y. Inoue},
\author[Physics]{M. Minowa}

\address[Physics]{Department of Physics, School of Science, University 
of Tokyo, 7-3-1, Hongo, Bunkyo-ku, Tokyo 113-0033, Japan}

\address[ICEPP]{International Center for Elementary Particle Physics(ICEPP), University of Tokyo, 7-3-1 Hongo, Bunkyo-ku, Tokyo 113-0033, Japan}

\corauth[cor1]{Corresponding author.
Tel.: +81 3 5841 7622; fax: +81 3 5841 4186.}

\begin{keyword}
Quadrupole Mass Spectrometer\sep Helium atoms \sep Counting
\end{keyword}

\begin{abstract}

We propose a new method of alpha($\alpha$)-ray measurement that detects helium
atoms with a Quadrupole Mass Spectrometer(QMS).  A demonstration is undertaken
with a plastic-covered $^{241}$Am $\alpha$-emitting source to detect
$\alpha$-rays stopped in the capsule.  We successfully detect helium
atoms that diffuse out of the capsule by accumulating them for one to 20
hours in a closed chamber. The detected amount is found to be proportional to 
the accumulation time. Our method is applicable to probe $\alpha$-emitting 
radioactivity in bulk material.

\end{abstract}

\end{frontmatter}

\section{Introduction}
There are many ways to detect $\alpha$ rays such as gas-flow counters
and solid-state detectors\cite{Alpha}, but all detect the corresponding 
ionization signal by the incident $\alpha$ ray, instead of $^4{\rm He}$ itself. 
However, because $\alpha$ particles travel only a few centimeters in the air 
and can be easily stopped by a piece of thin foil or paper, it is harder to detect 
their rays than those of beta or gamma radiation.

In this work, we suggest a new method of $\alpha$ ray measurement that aims to  
detect $^4{\rm He}$ neutral atoms. 
If one wants to measure $\alpha$-emitting radioactivity in bulk material with 
ordinary detectors, one has to rely on $\alpha$ rays emitted from the thin surface 
of the material because of their short range. However, many materials diffuse stopped
$\alpha$ particles out of their surface in the form of neutral helium atoms. 
The amount of the released helium atoms can then be measured by a Quadrupole Mass
Spectrometer (QMS) in terms of the mass number $A=4$. Therefore, there is the advantage 
of being able to measure $\alpha$-emitting radioactivity in bulk materials.
It can be seen that a higher sensitivity for alphas can be expected.  

\section{Experimental setup and method}
\begin{figure}[H]
\begin{center}
\includegraphics[width=4cm]{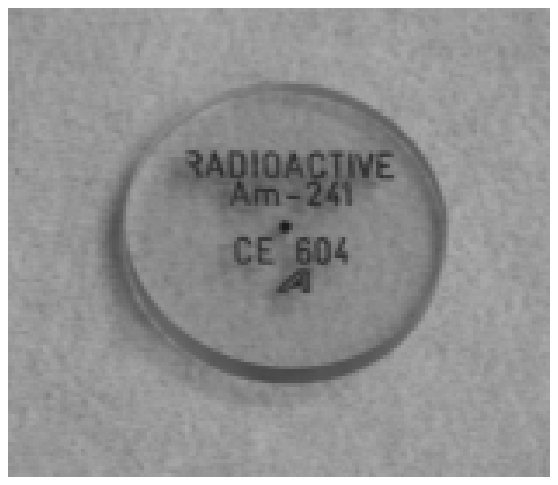}
\caption{Plastic-covered $^{241}{\rm Am}$ $\alpha$ source. The central ``point'' is $^{241}{\rm Am}$.}
\label{Am241-view}
\end{center}
\end{figure}

To examine our method of $\alpha$-ray measurement, we used an Amersham X.825 disc type
$^{241}{\rm Am}$ $\alpha$ source (Fig. \ref{Am241-view}) and a vacuum system including
a QMS. The $^{241}{\rm Am}$ source, an $\alpha$-emitter with a half life of 
$T_{1/2}=432.2$ y, is covered with epoxy resin 25mm in diameter and 3mm in thickness. 
This source emits $3.25\times10^5$ alphas per second, which eventually stop in the resin 
and diffuse out as $^4{\rm He}$ neutral atoms. The time needed for diffusion in resin is 
negligible\footnote{Diffusion time of helium through various samples of epoxy resin 
adhesives is found in ref \cite{DT}.}
compared to the time elapsed of about 10 years or more after the production of this 
$\alpha$ source. Therefore, the helium production rate of the source is 
thought to be the same as that of alpha ($3.25\times10^5$ s$^{-1}$).
A Quadrupole Mass Spectrometer (QMS) is often used for trace element 
analysis because of its high isotopic selectivity and efficiency\cite{QMS-ref}.
The QMS we used is a Pfeiffer Vacuum QMS200 with a channeltron detector. The mass range 
is $A=\mbox{1--100}$ and the detection limit is $1\times10^{-12}$ Pa \cite{Pfeiffer}.

\begin{figure}[H]
\begin{center}
\includegraphics[width=12cm]{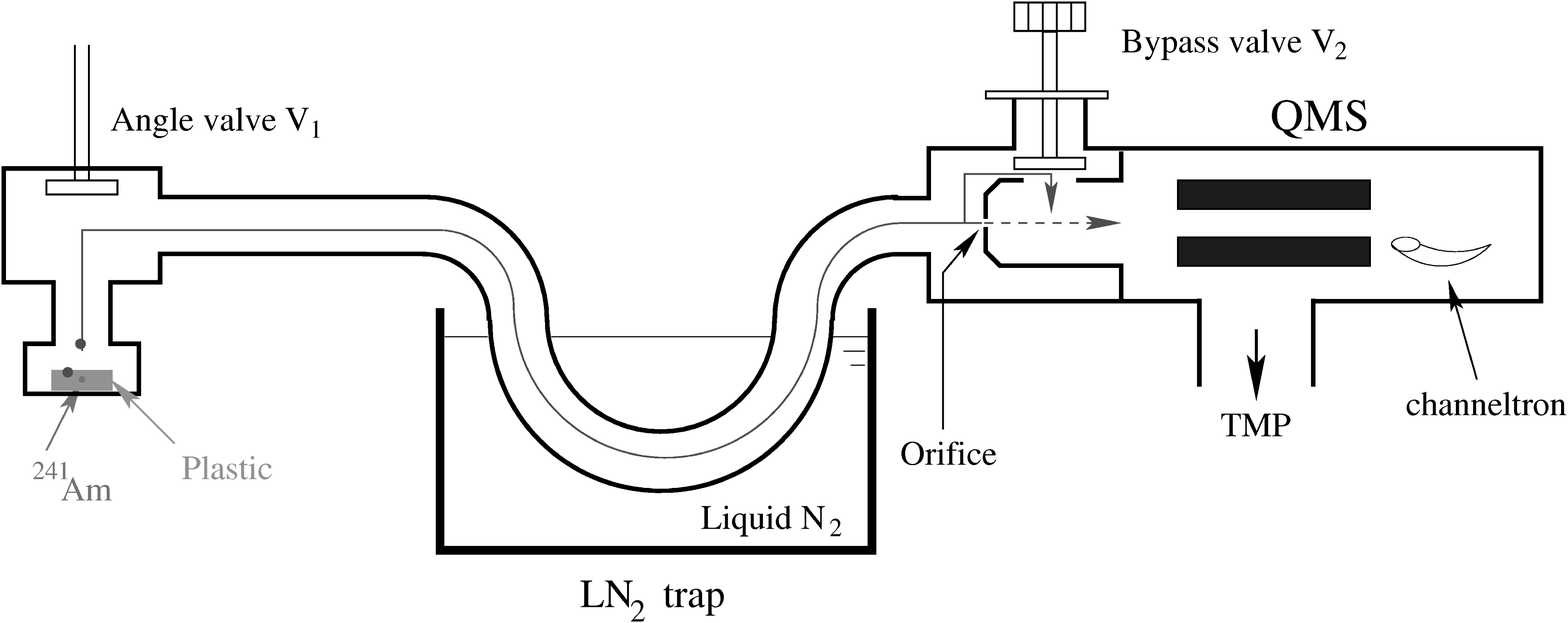}
\caption{Schematic view of our detection method.}
\label{sche-view}
\end{center}
\end{figure}

\begin{figure}[H]
\begin{center}
\includegraphics[width=10cm]{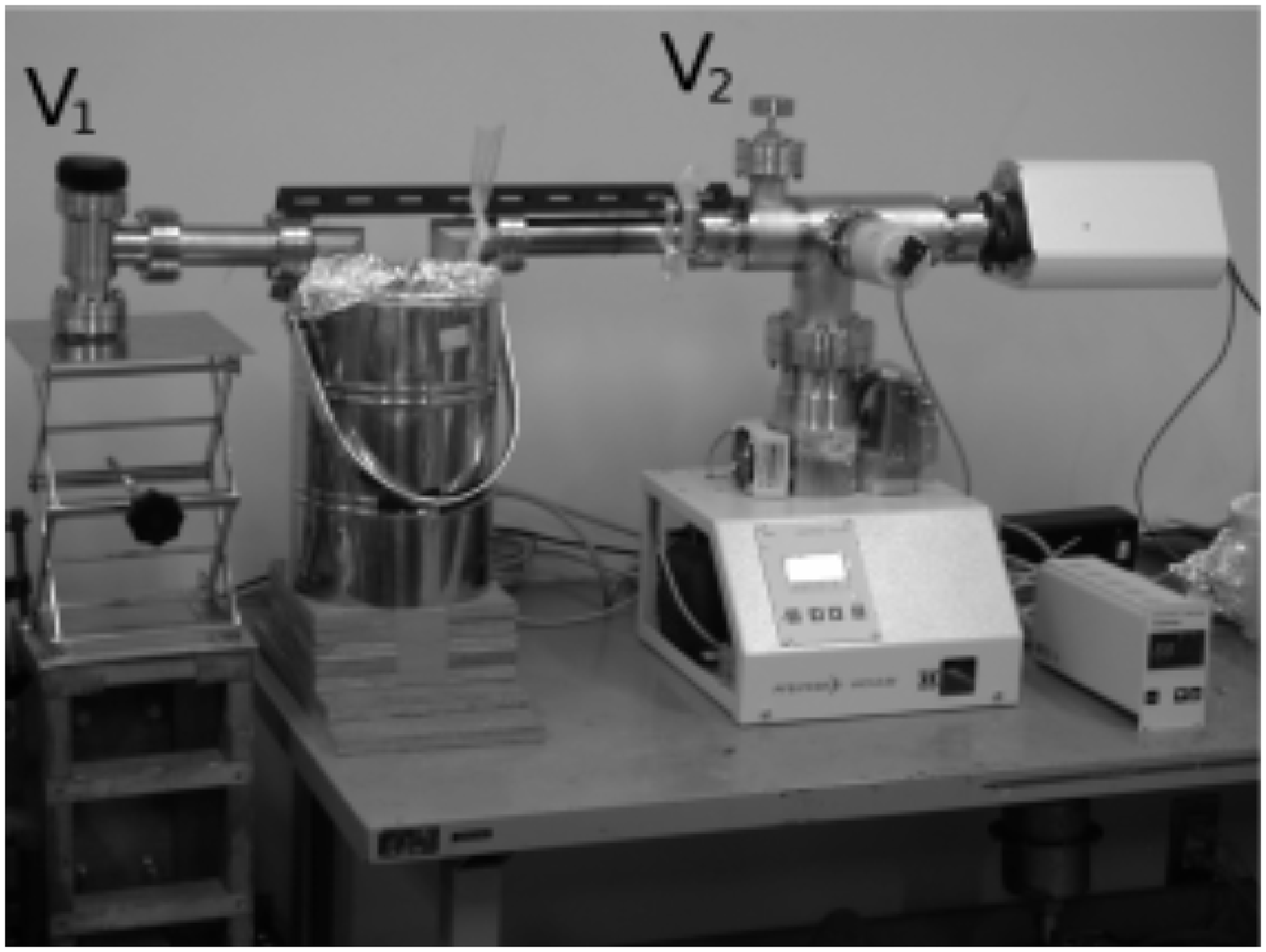}
\caption{Photograph of the experimental setup.}
\label{photo-view}
\end{center}
\end{figure}

The experimental setup is shown in Figs. \ref{sche-view} and \ref{photo-view}. 
A liquid nitrogen trap was introduced to capture unwanted out-gas especially
from the resin.
We controlled the valves, $V_1$ and $V_2$ (see Fig. \ref{photo-view}), to measure 
the integrated QMS channeltron current $Q_{\rm He}$ of helium as follows:
\begin{enumerate}
\item[(a)] Draw a vacuum to $\sim 10^{-6}$Pa with both valves ($V_1,V_2$) open.
\item[(b)] Keep valve $V_1$ closed for $T_{\rm ac}=$1h,4h,10h and 20h to 
accumulate helium atoms from the source. The number of helium atoms expected was 
$N_{\rm He}=1.17\times10^{9},\,4.68\times10^9,\,1.17\times10^{10}$ and 
$2.34\times10^{10}$, respectively.
\item[(c)] After closing valve $V_2$, open valve $V_1$ to introduce helium
atoms to the QMS through an orifice for a couple of seconds.
\item[(d)] Open valve $V_2$ to introduce most of the remaining atoms at one time
for helium detection.
\end{enumerate}
In our QMS system, the bypass valve $V_2$ and the orifice($\phi$0.3mm) are 
placed for a high pressure sample gas.
Here, we utilized them in step (c) to avoid a drastic change in the channeltron 
current measured by the QMS, which was observed with just $V_1$ open after step (b).
Considering the conductance of the orifice, the amount of helium atoms lost 
in step (c) is estimated to be less than 2$\%$, so we measured the integration 
of the channeltron currents for 5 seconds after opening valve $V_2$ in step (d). 
The integration time of 5s was determined after considering the time needed for opening 
valve $V_2$ by hand ($\sim1{\rm s}$) and the evacuation time $T_0\sim0.3$s ($\ll$5s) 
expected by the pumping speed of our vacuum system.

To estimate the amount of spurious signal caused by the out-gas 
from the resin, we also measured channeltron currents against a $^{57}{\rm Co}$ 
source as a control covered with the same epoxy capsule as the 
$^{241}{\rm Am}$ source.
Assuming that both sources are emitting out-gas of similar composition and 
amount, the net integrated channeltron current $Q_{\rm He}$ can be defined
as the difference between the integrated current $Q_{\rm Am}$ and $Q_{\rm Co}$ 
with the $^{241}{\rm Am}$ and the $^{57}{\rm Co}$ source, respectively, under the 
same accumulation time $T_{\rm ac}$. In terms of $N_{\rm He}$ described above, 
$Q_{\rm He}$ can be written as
\begin{equation}
Q_{\rm He}=Q_{\rm Am}-Q_{\rm Co}=eGR_iN_{\rm He},
\label{Q-He}
\end{equation}
where $e$ is the elementary charge, $G\sim7\times10^3$ is the amplification 
factor of the channeltron, and $R_i$ is the detection efficiency of the QMS, 
i.e., the ratio of the number of helium atoms detected by the QMS compared 
to the initial quantity $N_{\rm He}$. Note that $N_{\rm He}$ and $Q_{\rm He}$ are 
proportional to the accumulation time $T_{\rm ac}$.

\section{Result and analysis}
\begin{figure}[H]
\begin{center}
\begin{tabular}[b]{c}
\subfigure[$T_{\rm ac}=20$h.]{\includegraphics[width=9cm]{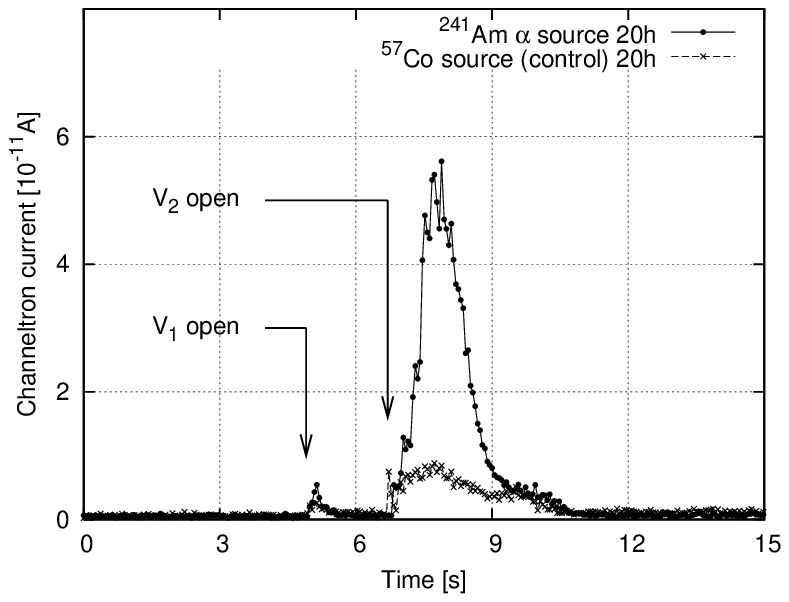}}\\
\subfigure[$T_{\rm ac}=10$h.]{\includegraphics[width=9cm]{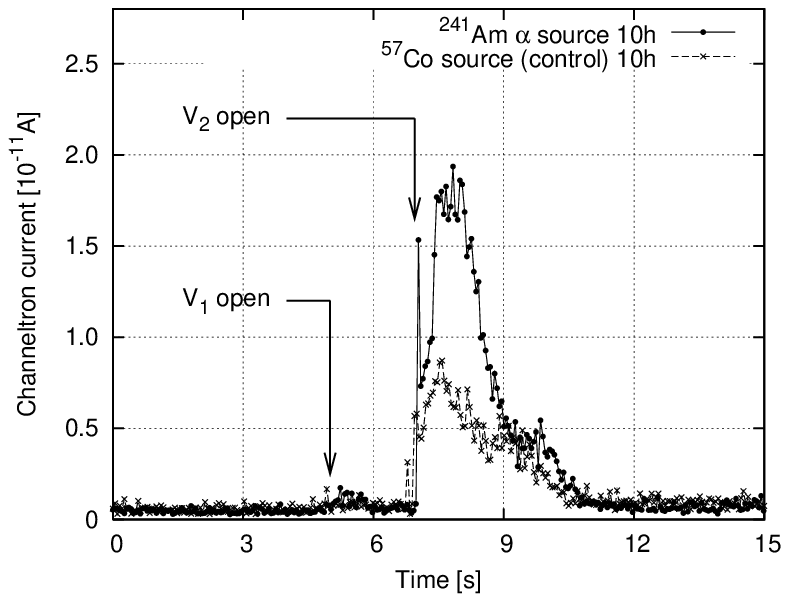}}\\
\subfigure[$T_{\rm ac}=4$h.]{\includegraphics[width=9cm]{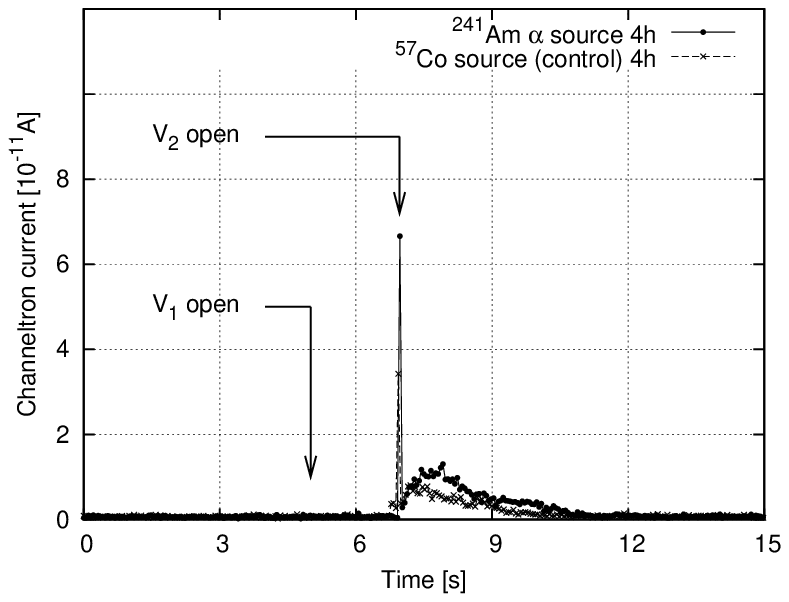}}
\end{tabular}
\caption{A few examples of the experimental result.}
\label{result-example}
\end{center}
\end{figure}

We measured a set of $Q_{\rm Am}$ and $Q_{\rm Co}$ twice for each 
$T_{\rm ac}=$1h,4h,10h and 20h. Fig. \ref{result-example} shows a few examples of 
the experimental result for $T_{\rm ac}=$ 20h, 10h and 4h. A clear difference between 
two sources can be seen for the data of $T_{\rm ac}=$ 20h and 10h. The difference can 
also be seen in Fig. \ref{result-example}-(C) ($T_{\rm ac}=$ 4h), though a drastic change 
in the channeltron current makes it unclear.

\begin{figure}[H]
\begin{center}
\includegraphics[width=12cm]{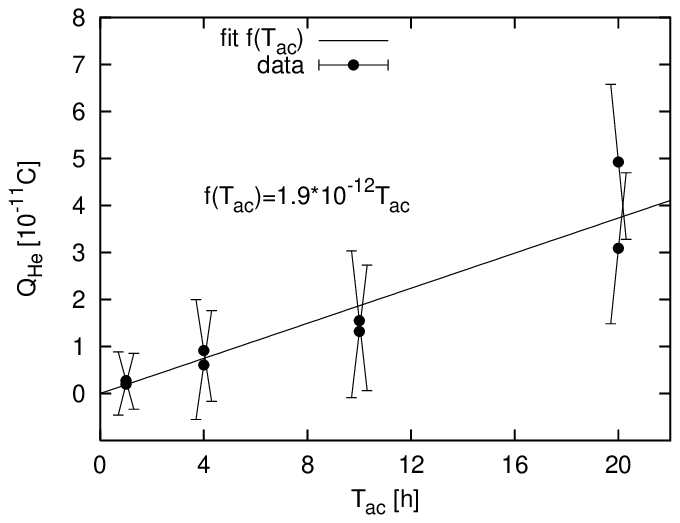}
\caption{Linearity between $T_{\rm ac}$ and $Q_{\rm He}$ ($T_{\rm ac}={\rm 1h,4h,10h,20h}$).}
\label{prop}
\end{center}
\end{figure}

Fig. \ref{prop} shows the relationship between the accumulation time $T_{\rm ac}$ 
and the net amount of helium $Q_{\rm He}=Q_{\rm Am}-Q_{\rm Co}$ with the best fit 
under the assumption of linearity. 
Here, because of the difficulty in evaluating the exact error of $Q_{\rm He}$, 
we estimated it at $Q_{\rm Co}$  for each data point, which may be a conservative 
overestimation. The unclarity for the data of $T_{\rm ac}=$ 4h described above is 
reflected in the error bars in this figure.

The best-fit proportionality coefficient 
$(1.9{\pm0.2})\times10^{-12}$[C/h] corresponds to $R_i\sim2\times10^{-6}$ according to 
Eq.(\ref{Q-He}). We can see certain degree of linearity between $T_{\rm ac}$ and 
$Q_{\rm He}$, but the obtained detection efficiency $R_i$ was rather small, probably 
because most of the helium atoms were evacuated by the vacuum pump before they got ionized. 

\begin{figure}[H]
\begin{center}
\includegraphics[width=12cm]{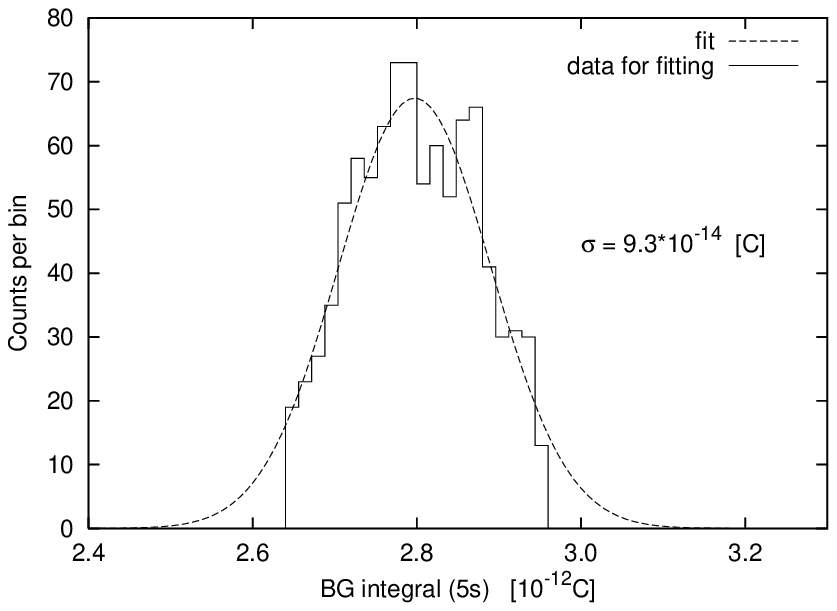}
\caption{Distribution of $Q_{\rm BG}$.}
\label{BG}
\end{center}
\end{figure}

To estimate the ultimate detection limit under this system apart from the influence 
of the out-gas coming from the epoxy capsule, we measured the distribution of 
$Q_{\rm BG}$, the background channeltron current without any sources 
integrated for 5 seconds, as shown in Fig. \ref{BG}. The best fit to the data by the 
normal distribution is also shown in the figure. In this way, the standard 
deviation $\sigma=9.3\times10^{-14}$ [C] was obtained. We defined the 
detection limit $Q_{\rm limit}$ as 
\begin{equation}
Q_{\rm limit}=(1.645\sigma)\times\sqrt{\mathstrut 2}=2.2\times10^{-13}
\hspace{0.2cm}{\rm [C]}
\end{equation}
at 95\% confidence level. Here, the factor $\sqrt{\mathstrut 2}$ is introduced 
to take into account the two sets of the data, the $^{241}{\rm Am}$ 
and the $^{57}{\rm Co}$ source, that were used for detecting the helium signal. 
The obtained $Q_{\rm limit}$ corresponds to $\sim10^8$ helium atoms under 
our system. 

Although we did not try to obtain higher detection efficiency 
$R_i$ with this measurement, since the valves were controlled manually, 
there are some possible solutions to improve $R_i$:

\begin{itemize}

\item Atom buncher \cite{buncher}\\
An atom buncher is a cold trap to capture target gas atoms.
It consists of a metal surface that often is cooled by liquid helium. 
Some kinds of laser can be used to temporarily heat a spot on the surface to 
evaporate gas atoms. 
The detection efficiency $R_i$ would improve if 
the heated area is placed close to the ionization chamber of the QMS.
This device can be applied to the detection of helium by cooling the surface
to a temperature low enough to trap the helium atoms.

\item Pulsed Supersonic Valve \cite{PSV}\\
The Pulsed Supersonic Valve(PSV) is an electromagnetic device to 
generate supersonic free gas jet.
This valve consists of two parallel metallic plates as a gate for the
sample gas. The gas is only allowed to be temporarily introduced when
the gate is opened by electromagnetic repulsion between these two plates. 
The jet of sample gas is then injected into the ionization chamber.
Similar to the atom buncher described above, higher $R_i$ can be achieved
by locating the PSV near the ionization region of the QMS.

\end{itemize}

Attaching devices such as an atom buncher and a PSV to improve 
detection efficiency and so obtain a higher $R_i$, much smaller quantities of 
helium atoms could be detected than under our present system. Also, some QMS 
are said to be able to count target atoms one by one\cite{Pfeiffer}, so our system 
could theoretically be improved to detect a single helium atom; that is, 
each $\alpha$ ray regardless of its energy if the detection efficiency $R_i$ 
gets close to one.

Our method can be applied to the accurate estimations of $\alpha$-induced soft errors 
in very-large-scale integrated circuit (VLSI). Soft errors are caused by 
$\alpha$ rays from a minute amount of radioactive substance in LSI packages, and 
have become a serious problem in VLSI circuits\cite{SE}. Using mass analysis in
vacuum may enable us to reduce the detection limit of $\alpha$-emitting radioactivity 
in the packages compared to the conventional method with a gas flow proportional counter.

\section{Conclusion}
We proposed a new method for alpha($\alpha$) ray measurement by detecting
helium atoms with a QMS. An $^{241}{\rm Am}$ $\alpha$ source and a 
$^{57}{\rm Co}$ source as a control were used to examine our method. 
The result showed that we could successfully detect helium atoms, but 
the detection efficiency was only $2\times10^{-6}$ under our system. 
Additional devices such as an atom buncher and a PSV to improve detection 
efficiency may allow us to reduce the detection limit.
This detector cannot measure the energy of the $\alpha$ particles, but 
that feature is reasonably common to many other conventional $\alpha$ 
detectors and is not a shortcoming in most applications.
Our method may become practicable to choose low $\alpha$-active material for LSI 
packages, which is essential to reduce $\alpha$-induced soft error rates in VLSI 
circuits.

\end{document}